\documentclass[a4paper,fleqn,usenatbib]{mnras}

\usepackage{newtxtext,newtxmath}

\usepackage[T1]{fontenc}
\usepackage{ae,aecompl}

\usepackage{graphicx}	
\usepackage{amsmath}	
\usepackage{amssymb}	

\usepackage{booktabs}

\usepackage{siunitx}
\DeclareSIUnit\Rjup{\ensuremath{\mathit{R}}_\mathrm{J}}
\DeclareSIUnit\Mjup{\ensuremath{\mathit{M}}_\mathrm{J}}

\DeclareSIUnit\Rsun{\ensuremath{\mathit{R}_{\sun}}}
\DeclareSIUnit\Msun{\ensuremath{\mathit{M}_{\sun}}}
\DeclareSIUnit\rhosun{\ensuremath{\rho_{\sun}}}

\DeclareSIUnit\Mearth{\ensuremath{\mathit{M}_{\earth}}}
\DeclareSIUnit\Rearth{\ensuremath{\mathit{R}_{\earth}}}

\DeclareSIUnit\Rstar{\ensuremath{\mathit{R}_{\star}}}

\DeclareSIUnit\days{d}

\DeclareSIUnit\cgs{cgs }
\DeclareSIUnit\logg{\ensuremath{\log{(\si{\centi\metre\per\square\second})}}}
\DeclareSIUnit\dex{dex }
\DeclareSIUnit\hjdutc{\ensuremath{\mathrm{HJD}_\mathrm{UTC}} }
\DeclareSIUnit\bjdutc{\ensuremath{\mathrm{BJD}_\mathrm{UTC}} }
\DeclareSIUnit\bjdtdb{$\mathrm{BJD}_\mathrm{TDB}$ }
\DeclareSIUnit\au{AU }
\DeclareSIUnit\pc{pc }

\DeclareSIUnit\yr{year }
\DeclareSIUnit\Gyr{Gyr }
\DeclareSIUnit\Myr{Myr }

\newcommand{\textblue}[1]{\textcolor{blue}{#1}}
\newcommand\coralie{CORALIE }
\newcommand\harps{HARPS }
\newcommand\euler{\emph{Euler} }
\newcommand\trappist{TRAPPIST }
\newcommand\wasp{WASP }

\newcommand*{\ra}[2][]{{
    \def\SIUnitSymbolDegree{\textsuperscript{h}}%
    \def\SIUnitSymbolArcminute{\textsuperscript{m}}%
    \def\SIUnitSymbolArcsecond{\textsuperscript{s}}%
    \ang[#1]{#2}}%
}

\newcommand\ve[3]{\ensuremath{#1^{+#2}_{-#3}}}
\newcommand\vep[3]{\ensuremath{#1\! \left(^{+#2}_{-#3}\right)}}

\newcommand\VDistance{422}
\newcommand\EDistance{6}

\newcommand\SpecType{G0V }

\newcommand\VTzero{2456720.68369}
\newcommand\EpTzero{19}
\newcommand\EmTzero{19}

\newcommand\VDepth{0.00699}
\newcommand\EpDepth{15}
\newcommand\EmDepth{15}

\newcommand\VWidth{0.11290}
\newcommand\EpWidth{48}
\newcommand\EmWidth{44}

\newcommand\VImpactParameter{0.11}
\newcommand\EpImpactParameter{10}
\newcommand\EmImpactParameter{07}

\newcommand\VSemiAmplitude{5.272}
\newcommand\EpSemiAmplitude{16}
\newcommand\EmSemiAmplitude{16}

\newcommand\VRadiusRatio{0.08359}
\newcommand\EpRadiusRatio{89}
\newcommand\EmRadiusRatio{88}

\newcommand\VMassRatio{0.03074}
\newcommand\EpMassRatio{37}
\newcommand\EmMassRatio{35}

\newcommand\VRadiusStar{1.152}
\newcommand\EpRadiusStar{21}
\newcommand\EmRadiusStar{16}

\newcommand\VRadiusScaledStar{0.1489}
\newcommand\EpRadiusScaledStar{24}
\newcommand\EmRadiusScaledStar{09}

\newcommand\VQOneR{0.3679}
\newcommand\EpQOneR{12}
\newcommand\EmQOneR{12}

\newcommand\VQTwoR{0.39437}
\newcommand\EpQTwoR{92}
\newcommand\EmQTwoR{92}

\newcommand\VQOneZ{0.3805}
\newcommand\EpQOneZ{12}
\newcommand\EmQOneZ{12}

\newcommand\VQTwoZ{0.39827}
\newcommand\EpQTwoZ{89}
\newcommand\EmQTwoZ{89}

\newcommand\VRadiusStarMean{1.16}
\newcommand\ERadiusStarMean{0.02}

\newcommand\VMassStar{1.155}
\newcommand\EpMassStar{39}
\newcommand\EmMassStar{39}

\newcommand\VMassStarMean{1.16}
\newcommand\EMassStarMean{0.04}

\newcommand\VloggStar{4.396}
\newcommand\EploggStar{08}
\newcommand\EmloggStar{14}

\newcommand\VDensityStar{0.807}
\newcommand\EpDensityStar{16}
\newcommand\EmDensityStar{38}

\newcommand\VSemiMajorAxis{0.03590}
\newcommand\EpSemiMajorAxis{38}
\newcommand\EmSemiMajorAxis{40}

\newcommand\VMassFunction{0.0000316}
\newcommand\EpMassFunction{03}
\newcommand\EmMassFunction{03}

\newcommand\VAgeMean{2.3}
\newcommand\EAgeMean{0.9}

\newcommand\VRadius{0.937}
\newcommand\EpRadius{22}
\newcommand\EmRadius{18}

\newcommand\VRadiusScaled{0.01246}
\newcommand\EpRadiusScaled{27}
\newcommand\EmRadiusScaled{18}

\newcommand\VRadiusMean{0.94}
\newcommand\ERadiusMean{0.02}

\newcommand\VMass{37.19}
\newcommand\EpMass{83}
\newcommand\EmMass{85}

\newcommand\VMassMean{37.5}
\newcommand\EMassMean{0.8}

\newcommand\VInclination{89.10}
\newcommand\EpInclination{63}
\newcommand\EmInclination{91}

\newcommand\VEccentricity{0.003}
\newcommand\EpEccentricity{0.003}
\newcommand\EmEccentricity{0.002}

\newcommand\Vlogg{5.040}
\newcommand\Eplogg{13}
\newcommand\Emlogg{18}

\newcommand\VDensity{55.9}
\newcommand\EpDensity{2.6}
\newcommand\EmDensity{3.5}

\usepackage{caption}
 \title[WASP-128b: a transiting
brown dwarf]{WASP-128b: a transiting brown dwarf in the \mbox{dynamical-tide} regime\thanks{using data collected at ESO's La Silla Observatory, Chile: HARPS on
        the ESO 3.6m (Prog IDs 095.C-0105 \& 097.C-0434), the Swiss {\it Euler}
    telescope, and TRAPPIST. The data is publicly available at CDS, and on
demand to the main author.}}

\author[V. Hod\v{z}i\'c et al.]{Vedad Hod\v{z}i\'c$^{\textblue{1}\thanks{E-mail:
    \href{mailto:vxh710@bham.ac.uk}{vxh710@bham.ac.uk}}}$,
    Amaury H.\,M.\,J. Triaud$^{\textblue{1}}$,
    David R. Anderson$^{\textblue{2}}$,
    Fran\c cois Bouchy$^{\textblue{3}}$,\newauthor
    Andrew Collier Cameron$^{\textblue{4}}$,
    Laetitia Delrez$^{\textblue{5}}$,
    Micha\"el Gillon$^{\textblue{6}}$,
    Coel Hellier$^{\textblue{2}}$,\newauthor
    Emmanu\"el Jehin$^{\textblue{6}}$,
    Monika Lendl$^{\textblue{7},\,\textblue{3}}$,
    Pierre F.\,L. Maxted$^{\textblue{2}}$,
    Francesco Pepe$^{\textblue{3}}$,\newauthor
    Don Pollacco$^{\textblue{8}}$,
    Didier Queloz$^{\textblue{5},\,\textblue{3}}$,
    Damien S\'egransan$^{\textblue{3}}$,
    Barry Smalley$^{\textblue{2}}$,\newauthor
    St\'ephane Udry$^{\textblue{3}}$,
    and Richard West$^{\textblue{8}}$
\\
\\
$^{1}$School of Physics and Astronomy, University of Birmingham, Edgbaston,
Birmingham B15 2TT, UK \\
$^{2}$Astrophysics Group, Keele University, Staffordshire, ST5 5BG, UK \\
$^{3}$Observatoire de Gen\`eve, Universit\'e de Gen\`eve, Chemin des Maillettes 51,
1290 Sauverny, Switzerland \\
$^{4}$Centre for Exoplanet Science, SUPA School of Physics and Astronomy, University of St. Andrews, North
Haugh, Fife, KY16 9SS, UK \\
$^{5}$Cavendish Laboratory, J\,J Thomson Avenue, Cambridge, CB3 0HE, UK \\
$^{6}$Institut d'Astrophysique et de G\'eophysique, Universit\'e de Li\`ege,
All\'ee du 6 Ao\^ut, 17, Bat. B5C, Li\`ege 1, Belgium \\
$^{7}$Space Research Institute, Austrian Academy of Sciences, Schmiedlstr. 6,
8042 Graz, Austria \\
$^{8}$Department of Physics, University of Warwick, Gibbet Hill Road, Coventry
CV4 7AL, UK \\
}

\date{Accepted 2018 August 28. Received 2018 August 27; in original form 2018
July 26}

\pubyear{2018}

\begin{document}
\label{firstpage}
\pagerange{\pageref{firstpage}--\pageref{lastpage}}
\maketitle
\sisetup{separate-uncertainty=true, multi-part-units=single}
\begin{abstract}
    Massive companions in close orbits around G
    dwarfs are thought to undergo rapid orbital decay due to runaway tidal 
    dissipation. We report here 
    the discovery of WASP-128b, a brown dwarf discovered by the WASP survey 
    transiting a \SpecType host on a $\SI{2.2}{\days}$ orbit, where the measured
    stellar rotation rate places the companion in a regime where tidal 
    interaction is dominated by dynamical tides. 
    Under the assumption of dynamical equilibrium, we derive a value of the
    stellar tidal quality factor $\log{Q_\star'} = \SI{6.96 \pm 0.19}{}$.
    A combined analysis of ground-based photometry and high-resolution
    spectroscopy reveals a mass and radius of the host, $M_\star = 
    \SI{\VMassStarMean \pm \EMassStarMean}{\Msun}$, $R_\star =
    \SI{\VRadiusStarMean \pm \ERadiusStarMean}{\Rsun}$, and for the companion, 
    $M_\mathrm{b} = \SI{\VMassMean \pm \EMassMean}{\Mjup}$, $R_\mathrm{b} =
    \SI{\VRadiusMean \pm \ERadiusMean}{\Rjup}$,
    placing WASP-128b in the driest parts of the brown dwarf desert, and
    suggesting a mild inflation for its age. 
    We estimate a remaining lifetime for WASP-128b similar to that of 
    some ultra-short period massive hot Jupiters, and note it may be a 
    propitious candidate for measuring orbital decay and testing tidal
    theories.
\end{abstract}




\section{Introduction}
Brown dwarfs are substellar objects that occupy the mass range 
${\sim}$\SIrange[range-phrase=--, range-units=single]{13}{80}{} Jupiter masses,
(\si{\Mjup}), thought to form via gravitational instability or molecular cloud 
fragmentation \citep{chabrier2014}.
Despite their
abundance, however, very little is known about brown dwarfs.
Most are found to be solitary, show complex spectral features that are difficult
to model, and their masses are typically hard to estimate because the models are degenerate
with their age, radius, and metallicity.
Brown dwarf companions orbiting Sun-like stars offer a chance to
study these objects in more detail as the stellar ages can be tied to the orbiting
brown dwarf. Moreover, transit light
curves can lift the inclination angle degeneracy to yield an unambiguous mass from
radial velocity measurements, providing precise physical parameters that
are crucial for testing substellar evolutionary models. 

\sisetup{range-phrase=--, range-units=single} 

Despite being fully sensitive throughout the brown dwarf mass range, early Doppler surveys
reported that brown dwarf companions are found in fewer numbers than their
free-floating counterparts, termed the \emph{brown dwarf desert} 
\citep{marcy_butler2000,sahlmann2011,ma_ge2014}. 
When comparing the same sample of host stars, up to \SI{16}{\percent} of
Sun-like stars have companions more massive than Jupiter, of which
${<}\SI{1}{\percent}$ are brown dwarfs \mbox{\citep{grether_lineweaver2006}}. 
Only twelve transiting brown dwarfs have been found to date
(\citealt{bayliss2017} and references therein, \citealt{canas2018}), 
where just three have been detected from the ground, possibly due to a detection
bias \citep{csizmadia2015}. 
Only one other brown dwarf has been 
discovered by the WASP survey (WASP-30b; \citealt{anderson2011,triaud2013}).

Most massive substellar companions on close orbits have been
found around F-type stars, and very few around G dwarfs, which has been
interpreted as being due to rapid engulfment of massive planets and brown dwarfs around G
dwarfs due to strong tidal coupling \citep{bouchy2011a,guillot2014,damiani2016}.
Stars generally spin down as they age due to magnetic braking,
where stellar winds carry highly ionised material that couples to the magnetic
field lines and gets carried away from the star, leading to angular
momentum loss. G dwarfs are typically more efficient at magnetic braking due to
their deeper outer convective layer.
However, companions on close orbits can transfer angular momentum
from the orbit to the stellar spin, thereby draining angular momentum from the
system via magnetic braking, leading to orbital decay until the companion is
engulfed by the host. 
The rate of the companion's orbital decay is predicted to increase by up to three orders of magnitude in
    the dynamical-tide regime \citep{ogilvie_lin2007}, with observational evidence
on hot Jupiter hosts supporting stronger tidal coupling than for equilibrium 
tides \citep{collier-cameron2018}. 
The strong dynamical tides from the companion excite inertial gravity
waves ($g$-modes) in the convective layer that, in G dwarfs, break and dissipate in the
radiative core, resulting in a spin-up the star from the inside \citep{barker_ogilvie2010,
essick2015}.
Thus in systems where the host stars can be spun up
by a massive companion such as a brown dwarf, the strong tidal coupling 
is expected to lead to runaway orbital decay of the companion onto the central star
on short timescales compared to the lifetime of the star \citep{barker_ogilvie2010}.

In this context we report the discovery of WASP-128b, a new transiting brown dwarf 
discovered by the WASP survey, orbiting
a \SpecType host on a close orbit, where the measured stellar rotation rate places the
system well-within the dynamical-tide regime, 
suggesting strong tidal coupling between the pair.

\section{Observations}
\subsection{Photometry}
WASP-128 is
a moderately bright ($V{=}12.5$) \SpecType star at a distance of
\SI[separate-uncertainty=true, multi-part-units=single]{422\pm5}{\pc}. The
\wasp survey \citep{pollacco2006} obtained \num{31543} images of WASP-128 between \mbox{2006-05-04}
and \mbox{2012-06-19}, identifying a periodic \SI{2.208}{\days} transit signal
in the photometry \citep{collier-cameron2006}. Consequently, we initiated photometric and spectroscopic 
follow-up observations.

Five transits of WASP-128b were obtained using the \SI{0.6}{\metre} \trappist robotic
telescope (\citealt{jehin2011,gillon2011}), located at ESO La Silla Observatory
(Chile). The first of these transit observations is partial, covering only the
second half of the transit. All
five transits were observed through a ``blue-blocking'' filter.
The images are calibrated using standard procedures (bias, dark, and
flat-field correction) and photometry is extracted using the
\textsc{iraf/daophot}\footnote{\textsc{iraf} is distributed by the National
Optical Astronomy Observatory, which is operated by the Association of
Universities for Research in Astronomy, Inc., under cooperative agreement with
the National Science Foundation.} aperture photometry software
(\citealt{stetson1987}), as described by \cite{gillon2013}.
For each transit
observation, a careful selection of both the photometric aperture size and
stable comparison stars is performed manually to obtain the most accurate
differential light curve of WASP-128.
Some light curves are affected by a
meridian flip; that is, the 180$^\circ$ rotation that TRAPPIST's 
equatorial mount has to undergo when the meridian is reached. We account for
any potential photometric offset in our baseline model, see Section~\ref{section:global_modelling}.

We observed three
transits of WASP-128b using the EulerCam instrument installed at the
\SI{1.2}{\metre} \euler
telescope also located at the La Silla site. The observations were carried out
through an $r'$-Gunn filter and the telescope was slightly defocused to improve
PSF sampling and observation efficiency. Each transit light curve is obtained
using relative aperture photometry while optimizing reference star selection and
extraction apertures to minimize the residual light curve RMS. The instrument
and the associated data reduction is described in more detail in
\citet{Lendl12}.
Some of the images of WASP-128 leading up to and during the ingress of the second 
transit were saturated, and as such, we discarded a handful of observations in
our analysis that had count levels above \num{50000} ADU.

\subsection{Spectroscopy}
We collected 48 spectra from the \coralie spectrograph on the 
\euler telescope between 2013-06-06 and 2016-11-24, as well as 23 HARPS spectra
on the ESO \SI{3.6}{\meter} telescope between 2015-04-02 and 2018-03-22.
Both sets of data are reduced using similar data reduction softwares. Their
resulting spectra are correlated with a numerical mask matching a G2V star
\mbox{\citep{baranne1996,pepe2002}}. These procedures have been demonstrated to reach
high precision and high accuracy \citep[e.g.][]{mayor09,lopezmorales2014}. We
perform a single $3\sigma$-clip on each radial velocity set using the line FWHM and 
bisector inverse slope span (BIS). One HARPS observation is discarded due to a highly
discrepant BIS value, and one CORALIE observation was
discarded due to the FWHM clip. Those outliers are highlighted in
Appendix~\ref{appendix:radial_velocity}.

\section{Data analysis}
\subsection{Spectral analysis}
\label{subsection:spectral_analysis} Using methods similar to those described by
\citet{doyle2013}, we used the co-added HARPS spectrum to determine
values for stellar effective temperature $T_\mathrm{eff}$, surface gravity $\log
g_\star$, metallicity [Fe/H], and projected stellar rotational velocity $v \sin
i_\star$. In determining $v \sin i$ we assumed a macroturbulent velocity of
4.4~$\pm$~0.7 km/s, based on the asteroseismic calibration of
\citet{doyle2014}. Using \textsc{mkclass} \citep{gray2014} we obtain a spectral
type \SpecType for WASP-128, which is consistent with the temperature derived
from the spectral analysis. The Lithium abundance $\log{A(\textrm{Li})} =
\SI{2.62 \pm 0.09}{}$ suggests a relatively young age of
\SIrange{1}{2}{\Gyr}.
\subsection{Global modelling} The combined data are analysed using
\label{section:global_modelling}
\texttt{amelie}, a novel software package that jointly models the photometric
and radial velocity data in a standard Bayesian framework. The code is
essentially a Python wrapper on the \texttt{ellc} binary star light curve model
\citep{maxted2016} for computing exoplanet and eclipsing binary light curves and
their radial velocity orbits, and the \texttt{emcee} affine-invariant Markov
chain Monte Carlo (MCMC) sampler \citep{goodman_weare2010, dfm2013} for
exploring the posterior parameter space.

We adopt a quadratic limb darkening law to model the intensity distribution of
the stellar disc, using the \texttt{LDTk} package \citep{husser2013, pyldtk} to
sample band-specific limb darkening coefficients for the \euler and \trappist
datasets, following the triangular parametrisation described in
\citet{kipping2013}.
No priors are imposed on the limb darkening parameters, rather, we fit the
intensity profile of the disc using the built-in likelihood function, allowing
uncertainties in the spectral parameters to be propagated to our final result.
In addition, we sample the following
parameters for our transit and radial velocity model: Period, $P$; mid-transit
reference time, $T_0$; transit depth, $D$; transit width (time from first to fourth
contact), $W$; impact parameter, $b$; and radial velocity semi-amplitude, $K$.
Moreover, for our eccentric model we also sample the parameters
$\sqrt{e}\sin{\omega}$ and $\sqrt{e}\cos{\omega}$, and in our orbital decay
model we further sample $\dot{P} = \mathrm{d}P/\mathrm{d}t$. 
In all cases we use non-informative priors that are either physically bounded 
(e.g. $0 < b < 1$) or 
sensibly bounded to a wide enough region 
(e.g. $0 < K < 50\,\si{\kilo\metre\per\second}$), where the transit midpoint is 
bounded by the light curves on 2014-03-03.

For each sampled set of parameters we further compute photometric and radial
velocity baseline models. Our photometric baseline model consists of a
normalisation factor with a second order polynomial in time for each light curve
to allow for airmass and seeing effects. Moreover, we experimented with
additional photometric detrending using sky background levels, FWHM changes in
the PSF, and changes in the target pixel position on the CCD.  Using the
Bayesian Information Criterion (BIC; \citealt{schwartz1978}) to compare model
complexity, we find that
an additional first-order polynomial using sky background levels is strongly
preferred for the \euler light curve on 2014-03-25.  On the nights of
2014-02-11, 2014-03-03, and 2015-05-10, the TRAPPIST telescope
peformed a meridian flip, for which we account for
any potential offsets by allowing an additional normalisation
factor before the flip.
The radial velocity baseline model consists of a constant systemic
velocity for each instrument. The \coralie data is partitioned into two
datasets
due to an upgrade of the instrument that could affect the velocity zero-point
\citep{triaud2017}.
We compare the constant velocity model with models allowing a first- and
second-order drift term, but find that any higher order terms are unjustified.
The baseline model parameters are computed using a least-squares
algorithm for every proposed parameter set in the MCMC
sampling. Finally, we also sample additional errors on our photometry and
radial velocity data to account for underestimated errors due to instrumental
effects and stellar activity.

The mean stellar density can be estimated independently from a transit light
curve and can be used with other observables to constrain the mass and age of a
star from stellar evolution models \citep{seager2003,triaud2013}.
We use \textsc{bagemass} \citep{maxted2015}
to estimate the age and mass of the host star, using our estimates 
of 
$T_\mathrm{eff}$ and [Fe/H] from the spectral modelling in
Section~\ref{subsection:spectral_analysis}, luminosity from \emph{Gaia}
DR2 \citep{gaiadr2}, and the mean stellar density from the transit light curves
as inputs to the code. The mass is then used as input to our Keplerian model. 

We initiate 256 walkers at positions normally dispersed close to
the solution, and run each walker for \num{30000} steps, chosen such that each
walker is run for a few tens of autocorrelation lengths after discarding the
first \num{15000} steps as burn-in. The independent
chains were thinned by a factor 100 due to autocorrelation, leaving each
parameter with \num{38400} independent samples, before 
computing the $\hat{R}$ statistic \citep{gelman2003}, 
and mixing the chains. 
All parameters
reach the recommended $\hat{R}{<}1.1$, indicating overall convergence.

\begin{figure}
    \includegraphics[width=\columnwidth]{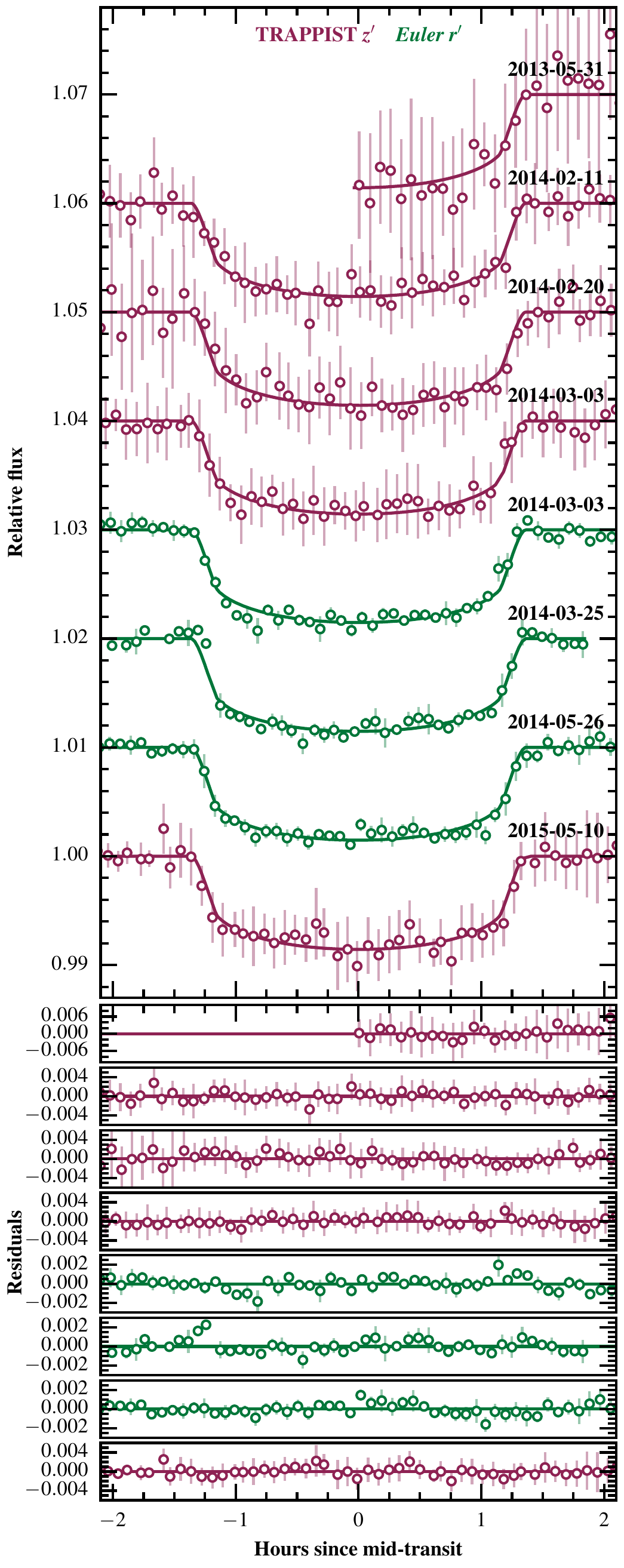}
    \caption{Transits of WASP-128 taken with the \euler (\emph{green}) and
    \trappist (\emph{vermillion}) telescopes. The points correspond to detrended
        data binned to 5 minutes, and the coloured lines are the best fit
        models. The residuals of the fit are shown in the lower panel.}
    \label{fig:data_lc}
\end{figure}
\begin{figure}
    \includegraphics[width=\columnwidth]{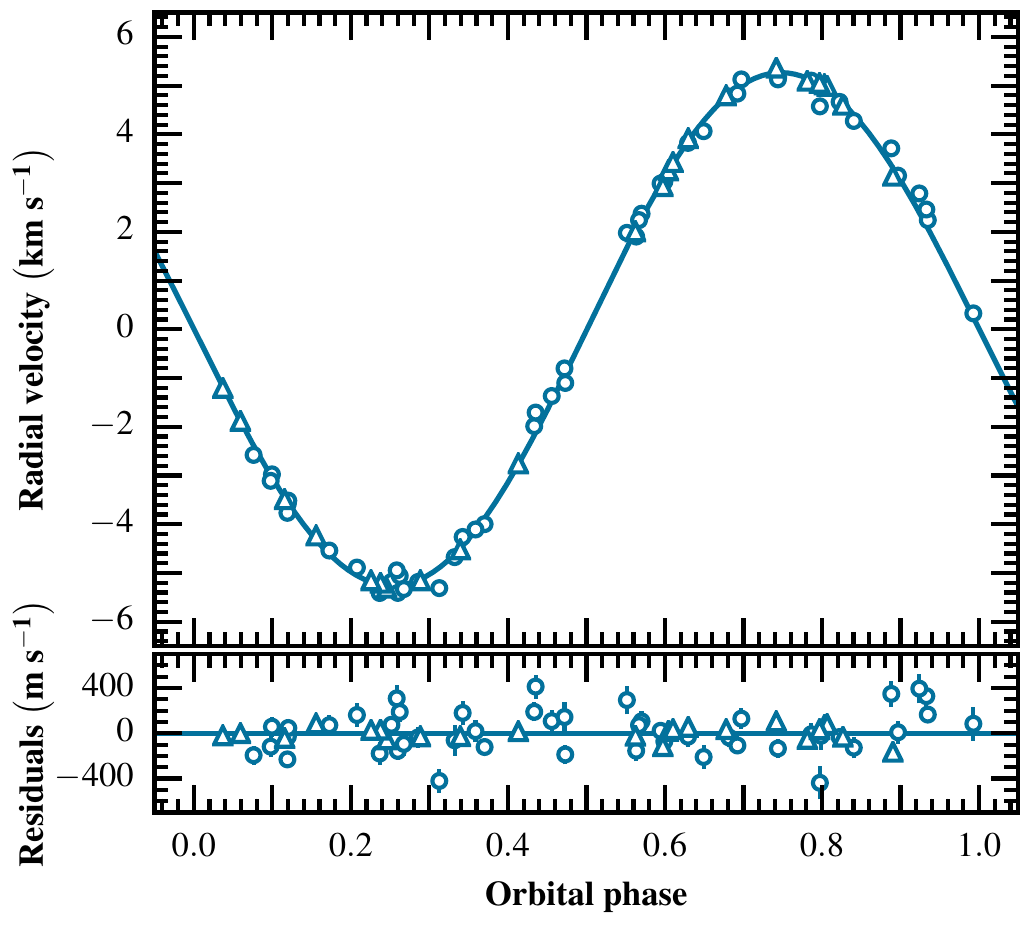}
    \caption{The radial velocity motion of WASP-128 due to its brown dwarf companion,
    folded on the best-fit period. The blue points correspond to RV measurements
    taken with \coralie (\emph{circles}) and \harps (\emph{triangles}). The solid
    line is the best-fit model. The residuals of the fit are shown in the lower
    panel.}
    \label{fig:data_rv}
\end{figure}

\section{Results}
\label{section:results}

\begin{table}
    \renewcommand{\arraystretch}{1.2}
    \newcolumntype{C}{ @{}>{${}}c<{{}$}@{} }
    \sisetup{separate-uncertainty}
    \sisetup{multi-part-units=single}
    \centering
    \caption{WASP128 system information and results. Numbers in brackets 
    denote uncertainties on the last two digits the 16th and 84th
    percentiles. $\star$ and ``b'' subscripts denote the host star and 
    companion, respectively.}
    \resizebox{\columnwidth}{!}{
                 \begin{tabular}{@{}l@{}
                     *1{rCl}@{}
                    s[table-unit-alignment = left]
                    l@{}
                    *1{rCl}@{}
                    s[table-unit-alignment = left]}
        \toprule
        \toprule
        \multicolumn{10}{@{}c@{}}{WASP-128} \\
        \multicolumn{10}{@{}c@{}}{\ra[angle-symbol-over-decimal]{11;31;26.10} 
        \ang[angle-symbol-over-decimal]{-41;41;22.3}} \\
        \multicolumn{10}{@{}c@{}}{2MASS J11312609-4141222} \\
        \multicolumn{10}{@{}c@{}}{\emph{Gaia} 5382697351745548416} \\
        \midrule
        \textbf{Parameter} & \multicolumn{3}{@{}c}{\textbf{Value}} & \textbf{Unit} &
        \textbf{Parameter} & \multicolumn{3}{@{}c}{\textbf{Value}} & \textbf{Unit} \\
        \midrule
        \multicolumn{10}{@{}l}{\emph{Spectral and system parameters}} \\
        $T_\mathrm{eff}$ & \multicolumn{3}{@{}r}{$5950\pm50$} & \si{\kelvin} & $d$ & 
        \multicolumn{3}{@{}r}{$\VDistance\pm\EDistance$} & \si{\pc} \\
        $\log{g_\star}$ & \multicolumn{3}{@{}r}{$4.1\pm0.1$} & \si{\cgs} & $\tau_\star$ &
        \multicolumn{3}{@{}r}{$2.2\pm0.9$} & \si{\Gyr} \\
        $[$Fe/H$]$ & \multicolumn{3}{@{}r}{$0.01\pm0.12$} & \si{\dex} & $G_\mathrm{mag}$ &
        \multicolumn{3}{@{}r}{$12.3$} \\
        $v\sin{i_\star}$ & \multicolumn{3}{@{}r}{$20.0\pm1.2$} & \si{\kilo\metre\per\second} &
        Sp.\,type & \multicolumn{3}{@{}r}{$\mathrm{G0V}$} \\
        \midrule
        \multicolumn{10}{@{}l}{\emph{Sampled parameters}} \\
        $P$ & \multicolumn{3}{@{}r}{\vep{2.208524}{21}{20}} & \si{\days} & $K$ &
        \multicolumn{3}{@{}r}{\vep{\VSemiAmplitude}{\EpSemiAmplitude}{\EmSemiAmplitude}}
        & \si{\kilo\metre\per\second} \\
%
        \multicolumn{4}{@{}r}{$T_0$ \hfill \vep{\num[group-digits=integer]{\VTzero}}{\EpTzero}{\EmTzero}} &
        \si{\bjdutc} &
        $q_1(r')$ & \multicolumn{3}{@{}r}{\vep{\VQOneR}{\EpQOneR}{\EmQOneR}} & \\
%
        $D$ & \multicolumn{3}{@{}r}{\vep{\VDepth}{\EpDepth}{\EmDepth}} & &
        $q_2(r')$ & \multicolumn{3}{@{}r}{\vep{\VQTwoR}{\EpQTwoR}{\EmQTwoR}} & \\
        $W$ & \multicolumn{3}{@{}r}{\vep{\VWidth}{\EpWidth}{\EmWidth}} & \si{\days} &
        $q_1(z')$ & \multicolumn{3}{@{}r}{\vep{\VQOneZ}{\EpQOneZ}{\EmQOneZ}} & \\
%
        $b$ &
        \multicolumn{3}{@{}r}{\vep{\VImpactParameter}{\EpImpactParameter}{\EmImpactParameter}}
        & \si{\Rstar} & 
        $q_2(z')$ & \multicolumn{3}{@{}r}{\vep{\VQTwoZ}{\EpQTwoZ}{\EmQTwoZ}} & \\
        \midrule
        \multicolumn{10}{@{}l}{\emph{Derived parameters}} \\
        $M_\star$ &
        \multicolumn{3}{@{}r}{\vep{\VMassStar}{\EpMassStar}{\EmMassStar}} & 
        \si{\Msun} & $M_\mathrm{b}$ &
        \multicolumn{3}{@{}r}{\vep{\VMass}{\EpMass}{\EmMass}} & \si{\Mjup} \\

        $R_\star$ & \multicolumn{3}{@{}r}{\vep{\VRadiusStar}{\EpRadiusStar}{\EmRadiusStar}}
        & \si{\Rsun} & $R_\mathrm{b}$ &
        \multicolumn{3}{@{}r}{\vep{\VRadius}{\EpRadius}{\EmRadius}} & \si{\Rjup} \\

        $R_\star/a$ &
        \multicolumn{3}{@{}r}{\vep{\VRadiusScaledStar}{\EpRadiusScaledStar}{\EmRadiusScaledStar}}
        & & $R_\mathrm{b}/a$ &
        \multicolumn{3}{@{}r}{\vep{\VRadiusScaled}{\EpRadiusScaled}{\EmRadiusScaled}} & \\

        $\rho_\star$ &
        \multicolumn{3}{@{}r}{\vep{\VDensityStar}{\EpDensityStar}{\EmDensityStar}}
        & \si{\rhosun} & $\rho_\mathrm{b}$ &
        \multicolumn{3}{@{}r}{\ve{\VDensity}{\EpDensity}{\EmDensity}}
        & \si{\gram\per\centi\metre\cubed} \\

        $\log{g_\star}$ &
        \multicolumn{3}{@{}r}{\vep{\VloggStar}{\EploggStar}{\EmloggStar}}
        & \si{\cgs} & $\log{g_\mathrm{b}}$ &
        \multicolumn{3}{@{}r}{\vep{\Vlogg}{\Eplogg}{\Emlogg}}
        & \si{\cgs} \\

        $a$ &
        \multicolumn{3}{@{}r}{\vep{\VSemiMajorAxis}{\EpSemiMajorAxis}{\EmSemiMajorAxis}}
        & \si{\au} & $i$ & 
        \multicolumn{3}{@{}r}{\vep{\VInclination}{\EpInclination}{\EmInclination}}
        & \si{\degree} \\

        $M_\mathrm{b}/M_\star$ &
        \multicolumn{3}{@{}r}{\vep{\VMassRatio}{\EpMassRatio}{\EmMassRatio}}
        & &
        $f(m)$ & 
        \multicolumn{3}{@{}r}{\vep{\VMassFunction}{\EpMassFunction}{\EmMassFunction}}
        & \si{\Msun} \\
        $R_\mathrm{b}/R_\star$ &
        \multicolumn{3}{@{}r}{\vep{\VRadiusRatio}{\EpRadiusRatio}{\EmRadiusRatio}}
        & & 
        $e$ & \multicolumn{3}{@{}r}{<\num{0.007}} & \\

        \bottomrule
    \end{tabular}}
    \label{tab:results}
\end{table}

\sisetup{separate-uncertainty=true, multi-part-units=single}
Using {\sc
bagemass} we find an age of \SI{\VAgeMean \pm \EAgeMean}{\Gyr} and mass of
$M_\star = \SI{\VMassStarMean \pm \EMassStarMean}{\Msun}$ for WASP-128. From
this we derive a radius of the star of $R_\star = \SI{\VRadiusStarMean \pm
\ERadiusStarMean}{\Rsun}$, and mass and radius of $M_\text{b}= \SI{\VMassMean \pm
\EMassMean}{\Mjup}$ and $R_\text{b} = \SI{\VRadiusMean \pm \ERadiusMean}{\Rjup}$,
for the companion, placing it securely in the brown dwarf regime. The best fit
models with the photometric data are shown in Fig.~\ref{fig:data_lc}, and in
Fig.~\ref{fig:data_rv} for the radial velocity data. The results from
our MCMC and derived parameters are shown in Table~\ref{tab:results} with their
\SI{68}{\percent} confidence interval.

\paragraph*{Eccentric model}{
Given the close proximity to the host
star, it is expected that the orbit of WASP-128b has been tidally circularised
due to tidal dissipation in the brown dwarf
as this would happen on a timescale of ${\leq}\SI{1}{\Gyr}$
\citep{barker_ogilvie2009}. Nevertheless, when including eccentricity in
our model, we derive a value of $e =
\ve{\VEccentricity}{\EpEccentricity}{\EmEccentricity}$.
Observational errors can lead to the detection of a small,
non-zero, but spurious eccentricity \citep{lucy_sweeney1971}.  The BIC strongly
disfavours an eccentric model compared to a circular fit. We apply the revised
Lucy-Sweeney test \citep{lucy2013} to put an upper limit of $e{<}0.007$ on the
eccentricity using their uniform prior. The results of the other parameters
between the two models are consistent with each other, and as such we present
the results from the circular fit in Table~\ref{tab:results},
with our upper limit on the eccentricity.}
\paragraph*{Orbital decay model}{ A periodic signal of \SI{2.93 \pm 0.03}{\days}
was found in the photometric data \citep{maxted2011}, which is consistent with
the derived rotation period of $P_\star = \SI{2.93 \pm 0.18}{\days}$ for
WASP-128, using the observed $v\sin{i_\star}$ and $R_\star$.  The fast rotation
for an early G dwarf could indicate a tidal spin-up due to its massive
companion. Given the expected strong tidal coupling in this system, we attempted
to directly measure an orbital decay in the radial velocity data by including a period time-derivative,
$\dot{P}$ in our model. We find that $\dot{P} =
1.05^{+1.13}_{-1.14}\,\si{\second\per\yr}$, and note that, although not 
significant, we would not expect a
positive $\dot{P}$ in our case where $P_\mathrm{orb} < P_\star$.

\section{Discussion}

\begin{figure}
    \includegraphics[width=\columnwidth]{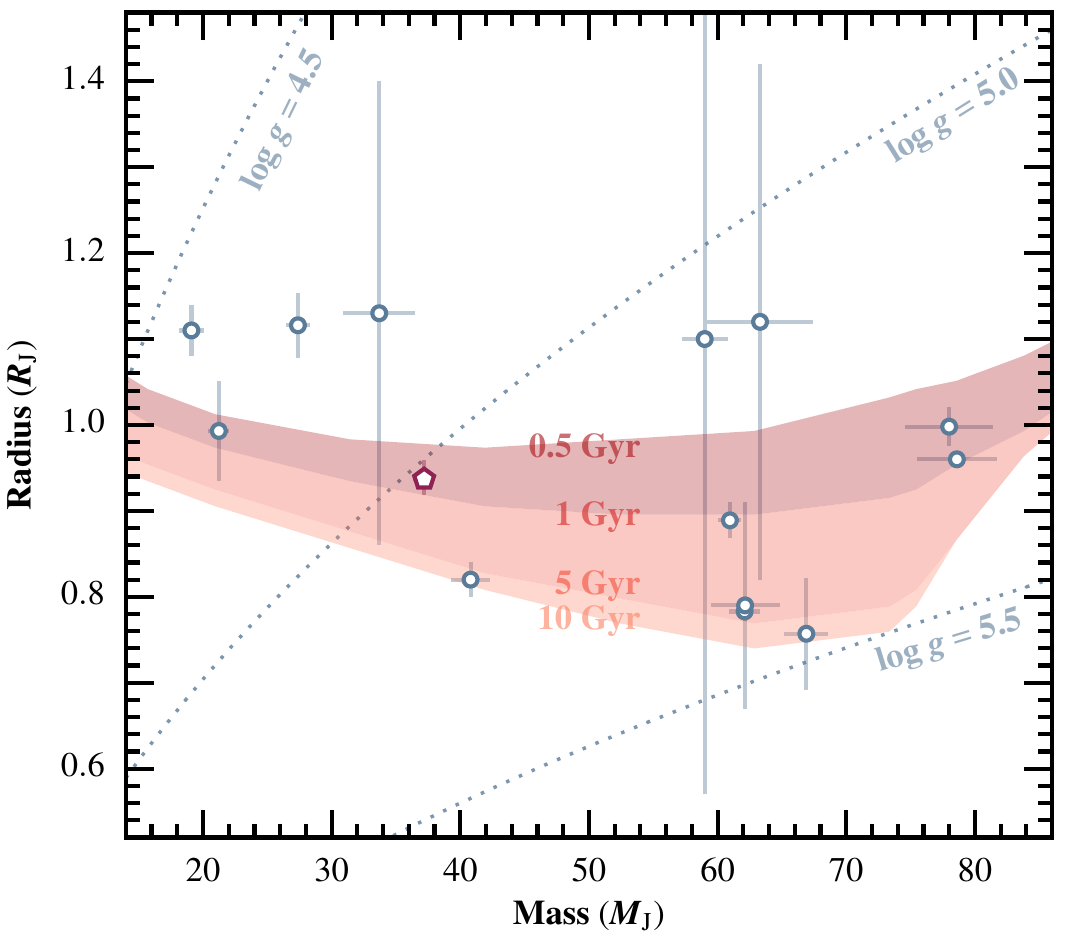}
    \caption{Mass-radius diagram placing WASP-128b (\emph{vermillion}) in the
    context of the other known transiting brown dwarfs. Objects are from
    \citet{bayliss2017} and references therein, including the recent discovery
    of Kepler-503b \citep{canas2018}. The shaded area outlines isochrones for
    substellar objects from the \texttt{COND03} models \citep{baraffe2003}.}
    \label{fig:mass-radius}
\end{figure}

\subsection{Tidal evolution}
\sisetup{separate-uncertainty=true, multi-part-units=single}
The orbital decay of the companion is significantly affected by magnetic braking
as long as the total angular momentum of the system can remain above the
critical limit \citep{damiani2015,damiani2016},
\begin{align*}
    L_\text{crit} &= 4\left( \frac{G^2}{27} \frac{M_\star^3 M_\text{b}^3}{M_\star +
        M_\text{b}} (\beta(t) I_\star + I_\text{b}) \right)^{\frac{1}{4}},
\end{align*}
where $I = \alpha MR^2$ is the moment of inertia of the two objects, defined by
their effective squared radii of gyration, $\alpha$. The inclusion of 
$\beta(t)$ extends the original work by \citet{hut1980} to include the
effects of orbital evolution due to magnetic braking, where $\beta = \Omega / n$
is the ratio of the stellar rotation rate to the orbital
frequency, with $n^2 = G(M_\star + M_\text{b})/a^3$. 
When $L > L_\text{crit}$, the system can enter a
pseudo-equilibrium state which is stable if the orbital angular momentum
$L_\text{orb}$ satisfies 
\begin{align*}
    L_\text{orb} > (4 - \beta(t)) (I_\star + I_\text{b})n.
\end{align*}
The companion satisfies both these criteria, so is thus either evolving towards the
dynamically stable state, or is already in synchronisation. Under the assumption
of the latter, we can derive the stellar tidal dissipation parameter $Q_\star'$
that is needed to balance the tidal torque with the wind braking torque using 
relations in e.g. \citet{brown2011} and \citet{damiani2016}, finding 
$\log{Q_\star'} = 6.96 \pm 0.19$.
Recently, \citet{collier-cameron2018} presented a study of the
hot Jupiter population that yielded a value of $\log{Q_\star'} = 
\SI{8.26 \pm 0.14}{}$.
In the regime where $0.5 < P/P_\star < 2$, dynamical tide become important \citep{ogilvie_lin2007}.
For a
subset of hot Jupiters that fall into this range, the tidal dissipation
parameter was found to be an order of magnitude smaller, where 
$\log{Q_\star'} = \SI{7.31 \pm 0.39}{}$, which is consistent with our estimate.
In fact, using the above estimate for $Q_\star'$, we derive that the spin period
of the star needed for a dynamically stable state is
$2.80_{-0.26}^{+0.44}$\,\si{\days}, which increases confidence in our
assumption about spin-orbit synchronisation. While in the dynamically stable state, 
the infall time of
WASP-128b is given by the magnetic braking timescale, and we derive a remaining
lifetime of $267^{+145}_{-67}\,\si{\Myr}$. In reality this is a lower limit, as
the infall time will not be driven by magnetic braking once the companion is
below the critical orbital period needed to stay in the dynamically stable
state. Thereafter, the infall will proceed more slowly, but will still reach the
Roche limit within a few
tens of \si{\Myr} (Fig. 3, \citealt{damiani2016}).

More generally, lifetime estimates depend on the structural and
rotational evolution of stars \citep{bolmont_mathis2016,gallet2017}.
Using $\log{Q_\star'} = 6$ and implementing a 
dynamical model that includes tidal interactions between the star and
companion, stellar evolution, magnetic braking, and tidal dissipation by gravity
waves, \citet{guillot2014} predicts a survival time of
\SIrange{50}{60}{\percent} of the host's main-sequence lifetime for a companion
at the mass of WASP-128b initially at a \SI{3}{\days} orbit. The main-sequence 
lifetime of
WASP-128 with a mass of about \SI{1.16}{\Msun} is ${\sim}\SI{6.9}{\Gyr}$, which
corresponds to a lifetime of \SIrange{3.5}{4.2}{\Gyr} for WASP-128b. The age
estimated from \textsc{bagemass} could thus be consistent with the companion's
survival, although a 
thorough calculation of the companion's evolutionary 
history is needed to estimate its initial location \citep{brown2011}.

\subsection{Inflation}
In Fig.~\ref{fig:mass-radius} we place WASP-128b in a mass-radius diagram with 
the other known transiting brown dwarfs. 
WASP-128b sits in the driest part of the brown dwarf desert, $35 < m\sin{i} <
55\,\SI{}{\Mjup}$ \citep{sahlmann2011,ma_ge2014}, coinciding with the 
${\sim}\SI{45}{\Mjup}$ mass minimum found in \citet{grether_lineweaver2006}. It
has been suggested that this minimum separates two brown dwarf populations 
differing by their formation mechanisms: The first formed 
in the protoplanetary disc via gravitational instability, and the second 
through molecular cloud fragmentation \citep{ma_ge2014}. In this context, 
WASP-128b clearly belongs to the low-mass population of brown dwarfs. 

Using our mass and age estimates for WASP-128b, the \texttt{COND03} 
evolutionary models \citep{baraffe2003} predict a radius of \SI{0.90}{\Rjup},
which suggests a mild inflation compared to the measured radius. Irradiation effects
should have little impact in inflating brown dwarfs, thus it is more likely due
to some other mechanism that deposits energy in the brown dwarf interior
\citep{bouchy2011a}. 

\section{Conclusion}
We have discovered WASP-128b, a transiting brown dwarf from the WASP survey on a
\SI{2.2}{\days} period around a G dwarf. Dynamical-tide theory predicts very few
such objects should exist due to rapid orbital decay from strong stellar 
tidal coupling. Using radial velocity data collected over ${\sim}5$ years, we
rule out any significant orbital decay, and we derive a value of the 
stellar tidal quality factor based on an assumption of dynamical stability. 
The derived age, mass, and size of
WASP-128b suggests a mild inflation, although we can not rule out a young
age.

\section*{Acknowledgements} The authors are grateful to the anonymous referee for
providing useful comments.
V.~H. is supported by the School of Physics \& Astronomy at
the University of Birmingham, as well a generous grant through the 
Postgraduate Research Scholarship Fund at the University of Birmingham.
We would like to thank the kind attention of the ESO
staff at La Silla, and the many observers who collected data with CORALIE and
HARPS along the years. 
This work has made use of data from the European
Space Agency (ESA) mission {\it Gaia} (\url{https://www.cosmos.esa.int/gaia}),
processed by the {\it Gaia} Data Processing and Analysis Consortium (DPAC,
\url{https://www.cosmos.esa.int/web/gaia/dpac/consortium}). Funding for the DPAC
has been provided by national institutions, in particular the institutions
participating in the {\it Gaia} Multilateral Agreement.  This research also made use
of Astropy, a community-developed core Python package for Astronomy
\citep{astropy}, as well as the open-source Python packages
Numpy \citep{numpy}, SciPy \citep{scipy}, and Matplotlib
\citep{matplotlib}.



\bibliographystyle{mnras}
\bibliography{references} 

\begin{thebibliography}{}
\makeatletter
\relax
\def\mn@urlcharsother{\let\do\@makeother \do\$\do\&\do\#\do\^\do\_\do\%\do\~}
\def\mn@doi{\begingroup\mn@urlcharsother \@ifnextchar [ {\mn@doi@}
  {\mn@doi@[]}}
\def\mn@doi@[#1]#2{\def\@tempa{#1}\ifx\@tempa\@empty \href
  {http://dx.doi.org/#2} {doi:#2}\else \href {http://dx.doi.org/#2} {#1}\fi
  \endgroup}
\def\mn@eprint#1#2{\mn@eprint@#1:#2::\@nil}
\def\mn@eprint@arXiv#1{\href {http://arxiv.org/abs/#1} {{\tt arXiv:#1}}}
\def\mn@eprint@dblp#1{\href {http://dblp.uni-trier.de/rec/bibtex/#1.xml}
  {dblp:#1}}
\def\mn@eprint@#1:#2:#3:#4\@nil{\def\@tempa {#1}\def\@tempb {#2}\def\@tempc
  {#3}\ifx \@tempc \@empty \let \@tempc \@tempb \let \@tempb \@tempa \fi \ifx
  \@tempb \@empty \def\@tempb {arXiv}\fi \@ifundefined
  {mn@eprint@\@tempb}{\@tempb:\@tempc}{\expandafter \expandafter \csname
  mn@eprint@\@tempb\endcsname \expandafter{\@tempc}}}

\bibitem[\protect\citeauthoryear{{Anderson} et~al.,}{{Anderson}
  et~al.}{2011}]{anderson2011}
{Anderson} D.~R.,  et~al., 2011, \mn@doi [\apjl] {10.1088/2041-8205/726/2/L19},
  \href {http://adsabs.harvard.edu/abs/2011ApJ...726L..19A} {726, L19}

\bibitem[\protect\citeauthoryear{{Andrae} et~al.,}{{Andrae}
  et~al.}{2018}]{gaiadr2}
{Andrae} R.,  et~al., 2018, preprint, \href
  {http://adsabs.harvard.edu/abs/2018arXiv180409374A} {} (\mn@eprint {arXiv}
  {1804.09374})

\bibitem[\protect\citeauthoryear{{Baraffe}, {Chabrier}, {Barman}, {Allard}  \&
  {Hauschildt}}{{Baraffe} et~al.}{2003}]{baraffe2003}
{Baraffe} I.,  {Chabrier} G.,  {Barman} T.~S.,  {Allard} F.,   {Hauschildt}
  P.~H.,  2003, \mn@doi [\aap] {10.1051/0004-6361:20030252}, \href
  {http://adsabs.harvard.edu/abs/2003A%26A...402..701B} {402, 701}

\bibitem[\protect\citeauthoryear{{Baranne} et~al.,}{{Baranne}
  et~al.}{1996}]{baranne1996}
{Baranne} A.,  et~al., 1996, \aaps, \href
  {http://adsabs.harvard.edu/abs/1996A%26AS..119..373B} {119, 373}

\bibitem[\protect\citeauthoryear{{Barker} \& {Ogilvie}}{{Barker} \&
  {Ogilvie}}{2009}]{barker_ogilvie2009}
{Barker} A.~J.,  {Ogilvie} G.~I.,  2009, \mn@doi [\mnras]
  {10.1111/j.1365-2966.2009.14694.x}, \href
  {https://ui.adsabs.harvard.edu/#abs/2009MNRAS.395.2268B} {395, 2268}

\bibitem[\protect\citeauthoryear{{Barker} \& {Ogilvie}}{{Barker} \&
  {Ogilvie}}{2010}]{barker_ogilvie2010}
{Barker} A.~J.,  {Ogilvie} G.~I.,  2010, \mn@doi [\mnras]
  {10.1111/j.1365-2966.2010.16400.x}, \href
  {https://ui.adsabs.harvard.edu/#abs/2010MNRAS.404.1849B} {404, 1849}

\bibitem[\protect\citeauthoryear{{Bayliss} et~al.,}{{Bayliss}
  et~al.}{2017}]{bayliss2017}
{Bayliss} D.,  et~al., 2017, \mn@doi [\aj] {10.3847/1538-3881/153/1/15}, \href
  {http://adsabs.harvard.edu/abs/2017AJ....153...15B} {153, 15}

\bibitem[\protect\citeauthoryear{{Bolmont} \& {Mathis}}{{Bolmont} \&
  {Mathis}}{2016}]{bolmont_mathis2016}
{Bolmont} E.,  {Mathis} S.,  2016, \mn@doi [Celestial Mechanics and Dynamical
  Astronomy] {10.1007/s10569-016-9690-3}, \href
  {http://adsabs.harvard.edu/abs/2016CeMDA.126..275B} {126, 275}

\bibitem[\protect\citeauthoryear{{Bouchy} et~al.,}{{Bouchy}
  et~al.}{2011}]{bouchy2011a}
{Bouchy} F.,  et~al., 2011, \mn@doi [\aap] {10.1051/0004-6361/201015276}, \href
  {http://adsabs.harvard.edu/abs/2011A%26A...525A..68B} {525, A68}

\bibitem[\protect\citeauthoryear{{Brown}, {Collier Cameron}, {Hall}, {Hebb}  \&
  {Smalley}}{{Brown} et~al.}{2011}]{brown2011}
{Brown} D.~J.~A.,  {Collier Cameron} A.,  {Hall} C.,  {Hebb} L.,   {Smalley}
  B.,  2011, \mn@doi [\mnras] {10.1111/j.1365-2966.2011.18729.x}, \href
  {http://adsabs.harvard.edu/abs/2011MNRAS.415..605B} {415, 605}

\bibitem[\protect\citeauthoryear{{Ca{\~n}as} et~al.,}{{Ca{\~n}as}
  et~al.}{2018}]{canas2018}
{Ca{\~n}as} C.~I.,  et~al., 2018, preprint, \href
  {http://adsabs.harvard.edu/abs/2018arXiv180508820C} {} (\mn@eprint {arXiv}
  {1805.08820})

\bibitem[\protect\citeauthoryear{{Chabrier}, {Johansen}, {Janson}  \&
  {Rafikov}}{{Chabrier} et~al.}{2014}]{chabrier2014}
{Chabrier} G.,  {Johansen} A.,  {Janson} M.,   {Rafikov} R.,  2014, \mn@doi
  [Protostars and Planets VI] {10.2458/azu_uapress_9780816531240-ch027}, \href
  {http://adsabs.harvard.edu/abs/2014prpl.conf..619C} {pp 619--642}

\bibitem[\protect\citeauthoryear{{Collier Cameron} \& {Jardine}}{{Collier
  Cameron} \& {Jardine}}{2018}]{collier-cameron2018}
{Collier Cameron} A.,  {Jardine} M.,  2018, \mn@doi [\mnras]
  {10.1093/mnras/sty292}, \href
  {http://adsabs.harvard.edu/abs/2018MNRAS.476.2542C} {476, 2542}

\bibitem[\protect\citeauthoryear{{Collier Cameron} et~al.,}{{Collier Cameron}
  et~al.}{2006}]{collier-cameron2006}
{Collier Cameron} A.,  et~al., 2006, \mn@doi [\mnras]
  {10.1111/j.1365-2966.2006.11074.x}, \href
  {http://adsabs.harvard.edu/abs/2006MNRAS.373..799C} {373, 799}

\bibitem[\protect\citeauthoryear{{Csizmadia} et~al.,}{{Csizmadia}
  et~al.}{2015}]{csizmadia2015}
{Csizmadia} S.,  et~al., 2015, \mn@doi [\aap] {10.1051/0004-6361/201526763},
  \href {https://ui.adsabs.harvard.edu/#abs/2015A&A...584A..13C} {584, A13}

\bibitem[\protect\citeauthoryear{{Damiani} \& {D{\'{\i}}az}}{{Damiani} \&
  {D{\'{\i}}az}}{2016}]{damiani2016}
{Damiani} C.,  {D{\'{\i}}az} R.~F.,  2016, \mn@doi [\aap]
  {10.1051/0004-6361/201527100}, \href
  {http://adsabs.harvard.edu/abs/2016A%26A...589A..55D} {589, A55}

\bibitem[\protect\citeauthoryear{{Damiani} \& {Lanza}}{{Damiani} \&
  {Lanza}}{2015}]{damiani2015}
{Damiani} C.,  {Lanza} A.~F.,  2015, \mn@doi [\aap]
  {10.1051/0004-6361/201424318}, \href
  {https://ui.adsabs.harvard.edu/#abs/2015A&A...574A..39D} {574, A39}

\bibitem[\protect\citeauthoryear{{Doyle} et~al.,}{{Doyle}
  et~al.}{2013}]{doyle2013}
{Doyle} A.~P.,  et~al., 2013, \mn@doi [\mnras] {10.1093/mnras/sts267}, \href
  {http://cdsads.u-strasbg.fr/abs/2013MNRAS.428.3164D} {428, 3164}

\bibitem[\protect\citeauthoryear{{Doyle}, {Davies}, {Smalley}, {Chaplin}  \&
  {Elsworth}}{{Doyle} et~al.}{2014}]{doyle2014}
{Doyle} A.~P.,  {Davies} G.~R.,  {Smalley} B.,  {Chaplin} W.~J.,   {Elsworth}
  Y.,  2014, \mn@doi [\mnras] {10.1093/mnras/stu1692}, \href
  {http://cdsads.u-strasbg.fr/abs/2014MNRAS.444.3592D} {444, 3592}

\bibitem[\protect\citeauthoryear{Essick \& Weinberg}{Essick \&
  Weinberg}{2016}]{essick2015}
Essick R.,  Weinberg N.~N.,  2016, The Astrophysical Journal, 816, 18

\bibitem[\protect\citeauthoryear{{Foreman-Mackey}, {Hogg}, {Lang}  \&
  {Goodman}}{{Foreman-Mackey} et~al.}{2013}]{dfm2013}
{Foreman-Mackey} D.,  {Hogg} D.~W.,  {Lang} D.,   {Goodman} J.,  2013, \mn@doi
  [\pasp] {10.1086/670067}, \href
  {http://adsabs.harvard.edu/abs/2013PASP..125..306F} {125, 306}

\bibitem[\protect\citeauthoryear{{Gallet}, {Bolmont}, {Mathis}, {Charbonnel}
  \& {Amard}}{{Gallet} et~al.}{2017}]{gallet2017}
{Gallet} F.,  {Bolmont} E.,  {Mathis} S.,  {Charbonnel} C.,   {Amard} L.,
  2017, \mn@doi [\aap] {10.1051/0004-6361/201730661}, \href
  {http://adsabs.harvard.edu/abs/2017A%26A...604A.112G} {604, A112}

\bibitem[\protect\citeauthoryear{Gelman, Carlin, Stern  \& Rubin}{Gelman
  et~al.}{2003}]{gelman2003}
Gelman A.,  Carlin J.,  Stern H.,   Rubin D.,  2003, Bayesian Data Analysis,
  Second Edition.
Chapman \& Hall/CRC Texts in Statistical Science, Taylor \& Francis, \url
  {https://books.google.co.uk/books?id=TNYhnkXQSjAC}

\bibitem[\protect\citeauthoryear{{Gillon}, {Jehin}, {Magain}, {Chantry},
  {Hutsem{\'e}kers}, {Manfroid}, {Queloz}  \& {Udry}}{{Gillon}
  et~al.}{2011}]{gillon2011}
{Gillon} M.,  {Jehin} E.,  {Magain} P.,  {Chantry} V.,  {Hutsem{\'e}kers} D.,
  {Manfroid} J.,  {Queloz} D.,   {Udry} S.,  2011, in European Physical Journal
  Web of Conferences. p. 06002 (\mn@eprint {arXiv} {1101.5807}),
  \mn@doi{10.1051/epjconf/20101106002}

\bibitem[\protect\citeauthoryear{{Gillon} et~al.,}{{Gillon}
  et~al.}{2013}]{gillon2013}
{Gillon} M.,  et~al., 2013, \mn@doi [\aap] {10.1051/0004-6361/201220561}, \href
  {http://adsabs.harvard.edu/abs/2013A%26A...552A..82G} {552, A82}

\bibitem[\protect\citeauthoryear{{Goodman} \& {Weare}}{{Goodman} \&
  {Weare}}{2010}]{goodman_weare2010}
{Goodman} J.,  {Weare} J.,  2010, \mn@doi [Communications in Applied
  Mathematics and Computational Science, Vol.~5, No.~1, p.~65-80, 2010]
  {10.2140/camcos.2010.5.65}, \href
  {http://adsabs.harvard.edu/abs/2010CAMCS...5...65G} {5, 65}

\bibitem[\protect\citeauthoryear{{Gray} \& {Corbally}}{{Gray} \&
  {Corbally}}{2014}]{gray2014}
{Gray} R.~O.,  {Corbally} C.~J.,  2014, \mn@doi [\aj]
  {10.1088/0004-6256/147/4/80}, \href
  {http://adsabs.harvard.edu/abs/2014AJ....147...80G} {147, 80}

\bibitem[\protect\citeauthoryear{{Grether} \& {Lineweaver}}{{Grether} \&
  {Lineweaver}}{2006}]{grether_lineweaver2006}
{Grether} D.,  {Lineweaver} C.~H.,  2006, \mn@doi [\apj] {10.1086/500161},
  \href {http://adsabs.harvard.edu/abs/2006ApJ...640.1051G} {640, 1051}

\bibitem[\protect\citeauthoryear{{Guillot}, {Lin}, {Morel}, {Havel}  \&
  {Parmentier}}{{Guillot} et~al.}{2014}]{guillot2014}
{Guillot} T.,  {Lin} D.~N.~C.,  {Morel} P.,  {Havel} M.,   {Parmentier} V.,
  2014, in EAS Publications Series. pp 327--336 (\mn@eprint {arXiv}
  {1409.7477}), \mn@doi{10.1051/eas/1465009}

\bibitem[\protect\citeauthoryear{Hunter}{Hunter}{2007}]{matplotlib}
Hunter J.~D.,  2007, \mn@doi [Computing In Science \& Engineering]
  {10.1109/MCSE.2007.55}, 9, 90

\bibitem[\protect\citeauthoryear{{Husser}, {Wende-von Berg}, {Dreizler},
  {Homeier}, {Reiners}, {Barman}  \& {Hauschildt}}{{Husser}
  et~al.}{2013}]{husser2013}
{Husser} T.-O.,  {Wende-von Berg} S.,  {Dreizler} S.,  {Homeier} D.,  {Reiners}
  A.,  {Barman} T.,   {Hauschildt} P.~H.,  2013, \mn@doi [\aap]
  {10.1051/0004-6361/201219058}, \href
  {http://adsabs.harvard.edu/abs/2013A%26A...553A...6H} {553, A6}

\bibitem[\protect\citeauthoryear{{Hut}}{{Hut}}{1980}]{hut1980}
{Hut} P.,  1980, \aap, \href
  {http://adsabs.harvard.edu/abs/1980A%26A....92..167H} {92, 167}

\bibitem[\protect\citeauthoryear{{Jehin} et~al.,}{{Jehin}
  et~al.}{2011}]{jehin2011}
{Jehin} E.,  et~al., 2011, The Messenger, \href
  {http://adsabs.harvard.edu/abs/2011Msngr.145....2J} {145, 2}

\bibitem[\protect\citeauthoryear{Jones, Oliphant, Peterson  et~al.}{Jones
  et~al.}{01  }]{scipy}
Jones E.,  Oliphant T.,  Peterson P.,   et~al., 2001--, {SciPy}: Open source
  scientific tools for {Python}, \url {http://www.scipy.org/}

\bibitem[\protect\citeauthoryear{{Kipping}}{{Kipping}}{2013}]{kipping2013}
{Kipping} D.~M.,  2013, \mn@doi [\mnras] {10.1093/mnras/stt1435}, \href
  {http://adsabs.harvard.edu/abs/2013MNRAS.435.2152K} {435, 2152}

\bibitem[\protect\citeauthoryear{{Lendl} et~al.,}{{Lendl}
  et~al.}{2012}]{Lendl12}
{Lendl} M.,  et~al., 2012, \mn@doi [\aap] {10.1051/0004-6361/201219585}, \href
  {http://adsabs.harvard.edu/abs/2012A%26A...544A..72L} {544, A72}

\bibitem[\protect\citeauthoryear{{L{\'o}pez-Morales}
  et~al.,}{{L{\'o}pez-Morales} et~al.}{2014}]{lopezmorales2014}
{L{\'o}pez-Morales} M.,  et~al., 2014, \mn@doi [\apjl]
  {10.1088/2041-8205/792/2/L31}, \href
  {http://adsabs.harvard.edu/abs/2014ApJ...792L..31L} {792, L31}

\bibitem[\protect\citeauthoryear{{Lucy}}{{Lucy}}{2013}]{lucy2013}
{Lucy} L.~B.,  2013, \mn@doi [\aap] {10.1051/0004-6361/201219819}, \href
  {https://ui.adsabs.harvard.edu/#abs/2013A&A...551A..47L} {551, A47}

\bibitem[\protect\citeauthoryear{{Lucy} \& {Sweeney}}{{Lucy} \&
  {Sweeney}}{1971}]{lucy_sweeney1971}
{Lucy} L.~B.,  {Sweeney} M.~A.,  1971, \mn@doi [\aj] {10.1086/111159}, \href
  {http://adsabs.harvard.edu/abs/1971AJ.....76..544L} {76, 544}

\bibitem[\protect\citeauthoryear{{Ma} \& {Ge}}{{Ma} \& {Ge}}{2014}]{ma_ge2014}
{Ma} B.,  {Ge} J.,  2014, \mn@doi [\mnras] {10.1093/mnras/stu134}, \href
  {http://adsabs.harvard.edu/abs/2014MNRAS.439.2781M} {439, 2781}

\bibitem[\protect\citeauthoryear{{Marcy} \& {Butler}}{{Marcy} \&
  {Butler}}{2000}]{marcy_butler2000}
{Marcy} G.~W.,  {Butler} R.~P.,  2000, \mn@doi [\pasp] {10.1086/316516}, \href
  {http://adsabs.harvard.edu/abs/2000PASP..112..137M} {112, 137}

\bibitem[\protect\citeauthoryear{{Maxted}}{{Maxted}}{2016}]{maxted2016}
{Maxted} P.~F.~L.,  2016, \mn@doi [\aap] {10.1051/0004-6361/201628579}, \href
  {http://adsabs.harvard.edu/abs/2016A%26A...591A.111M} {591, A111}

\bibitem[\protect\citeauthoryear{{Maxted} et~al.,}{{Maxted}
  et~al.}{2011}]{maxted2011}
{Maxted} P.~F.~L.,  et~al., 2011, \mn@doi [Publications of the Astronomical
  Society of the Pacific] {10.1086/660007}, \href
  {https://ui.adsabs.harvard.edu/#abs/2011PASP..123..547M} {123, 547}

\bibitem[\protect\citeauthoryear{{Maxted}, {Serenelli}  \&
  {Southworth}}{{Maxted} et~al.}{2015}]{maxted2015}
{Maxted} P.~F.~L.,  {Serenelli} A.~M.,   {Southworth} J.,  2015, \mn@doi [\aap]
  {10.1051/0004-6361/201425331}, \href
  {http://adsabs.harvard.edu/abs/2015A%26A...575A..36M} {575, A36}

\bibitem[\protect\citeauthoryear{{Mayor} et~al.,}{{Mayor}
  et~al.}{2009}]{mayor09}
{Mayor} M.,  et~al., 2009, \mn@doi [\aap] {10.1051/0004-6361:200810451}, \href
  {http://adsabs.harvard.edu/abs/2009A%26A...493..639M} {493, 639}

\bibitem[\protect\citeauthoryear{{Ogilvie} \& {Lin}}{{Ogilvie} \&
  {Lin}}{2007}]{ogilvie_lin2007}
{Ogilvie} G.~I.,  {Lin} D.~N.~C.,  2007, \mn@doi [\apj] {10.1086/515435}, \href
  {http://adsabs.harvard.edu/abs/2007ApJ...661.1180O} {661, 1180}

\bibitem[\protect\citeauthoryear{Parviainen \& Aigrain}{Parviainen \&
  Aigrain}{2015}]{pyldtk}
Parviainen H.,  Aigrain S.,  2015, \mn@doi [Monthly Notices of the Royal
  Astronomical Society] {10.1093/mnras/stv1857}, 453, 3821

\bibitem[\protect\citeauthoryear{{Pepe} et~al.,}{{Pepe}
  et~al.}{2002}]{pepe2002}
{Pepe} F.,  et~al., 2002, The Messenger, \href
  {http://adsabs.harvard.edu/abs/2002Msngr.110....9P} {110, 9}

\bibitem[\protect\citeauthoryear{{Pollacco} et~al.,}{{Pollacco}
  et~al.}{2006}]{pollacco2006}
{Pollacco} D.~L.,  et~al., 2006, \mn@doi [\pasp] {10.1086/508556}, \href
  {http://adsabs.harvard.edu/abs/2006PASP..118.1407P} {118, 1407}

\bibitem[\protect\citeauthoryear{{Sahlmann} et~al.,}{{Sahlmann}
  et~al.}{2011}]{sahlmann2011}
{Sahlmann} J.,  et~al., 2011, \mn@doi [\aap] {10.1051/0004-6361/201015427},
  \href {http://adsabs.harvard.edu/abs/2011A%26A...525A..95S} {525, A95}

\bibitem[\protect\citeauthoryear{{Schwarz}}{{Schwarz}}{1978}]{schwartz1978}
{Schwarz} G.,  1978, Annals of Statistics, \href
  {https://ui.adsabs.harvard.edu/#abs/1978AnSta...6..461S} {6, 461}

\bibitem[\protect\citeauthoryear{{Seager} \& {Mall{\'e}n-Ornelas}}{{Seager} \&
  {Mall{\'e}n-Ornelas}}{2003}]{seager2003}
{Seager} S.,  {Mall{\'e}n-Ornelas} G.,  2003, \mn@doi [\apj] {10.1086/346105},
  \href {http://adsabs.harvard.edu/abs/2003ApJ...585.1038S} {585, 1038}

\bibitem[\protect\citeauthoryear{{Stetson}}{{Stetson}}{1987}]{stetson1987}
{Stetson} P.~B.,  1987, \mn@doi [\pasp] {10.1086/131977}, \href
  {http://adsabs.harvard.edu/abs/1987PASP...99..191S} {99, 191}

\bibitem[\protect\citeauthoryear{{The Astropy Collaboration} et~al.,}{{The
  Astropy Collaboration} et~al.}{2018}]{astropy}
{The Astropy Collaboration} et~al., 2018, preprint, \href
  {http://adsabs.harvard.edu/abs/2018arXiv180102634T} {} (\mn@eprint {arXiv}
  {1801.02634})

\bibitem[\protect\citeauthoryear{{Triaud} et~al.,}{{Triaud}
  et~al.}{2013}]{triaud2013}
{Triaud} A.~H.~M.~J.,  et~al., 2013, \mn@doi [\aap]
  {10.1051/0004-6361/201219643}, \href
  {https://ui.adsabs.harvard.edu/#abs/2013A&A...549A..18T} {549, A18}

\bibitem[\protect\citeauthoryear{{Triaud} et~al.,}{{Triaud}
  et~al.}{2017}]{triaud2017}
{Triaud} A.~H.~M.~J.,  et~al., 2017, \mn@doi [\mnras] {10.1093/mnras/stx154},
  \href {http://adsabs.harvard.edu/abs/2017MNRAS.467.1714T} {467, 1714}

\bibitem[\protect\citeauthoryear{Walt, Colbert  \& Varoquaux}{Walt
  et~al.}{2011}]{numpy}
Walt S. v.~d.,  Colbert S.~C.,   Varoquaux G.,  2011, \mn@doi [Computing in
  Science and Engg.] {10.1109/MCSE.2011.37}, 13, 22

\makeatother
\end{thebibliography}




\appendix

\section{Radial velocities}
\label{appendix:radial_velocity}
\begin{table}
    \sisetup{round-mode=places}
    \centering
    \caption{HARPS radial velocity dataset. Machine readable format is available
    online at CDS.\newline
    $^*$marks data that was excluded from the fit.}
    \begin{tabular*}{\columnwidth}{@{\extracolsep{\fill}}
                    S[table-format=5.5, group-digits=false, 
                      round-precision=5, table-unit-alignment = center]
                    S[table-format=2.3,
                      round-precision=3, table-unit-alignment = center]
                    S[table-format=1.3,
                      round-precision=3, table-unit-alignment = center]
                    S[table-format=2.3,
                      round-precision=3, table-unit-alignment = center]
                    S[table-format=-3.3,
                      round-precision=3, table-unit-alignment = center]}
        \toprule
        \toprule
        {BJD$_\mathrm{UTC}$} & {RV} & {$\sigma$} & {FWHM} & {BIS} \\
         & {\SI{}{\kilo\metre\per\second}} &
        {\SI{}{\kilo\metre\per\second}} 
        & {\SI{}{\kilo\metre\per\second}} &
        {\SI{}{\kilo\metre\per\second}} \\
        \midrule
        57114.604490 &  10.19886 &   0.03857 &  28.02641 &  0.65954	  \\  
        57114.767777 &  11.96217 &   0.03481 &  28.04199 &  0.49807   \\
        57115.580569 &  19.81497 &   0.03248 &  28.21436 &  -0.43354  \\
        57115.821461 &  17.86571 &   0.03447 &  28.53128 &  -0.25040  \\
        57116.701126 &  9.56556  &   0.03357 &  27.96543 &  0.04673   \\
        57135.560277{$^*$} &  19.32195 &   0.04225 &  28.00931 &  -212.09835\\
        57137.581652 &  20.08288 &   0.03328 &  28.06999 &  -0.14360  \\
        57138.694412 &  9.41227  &   0.03810 &  27.94346 &  -0.07837  \\
        57139.649675 &  19.51470 &   0.03889 &  27.97368 &  -0.00470  \\
        57141.695657 &  17.97551 &   0.03256 &  28.06352 &  -0.23932  \\
        57157.596595 &  19.73995 &   0.03206 &  27.75519 &  -0.30482  \\
        57158.556607 &  9.51614  &   0.02926 &  27.90052 &  -0.15205  \\
        57181.510526 &  18.63475 &   0.03266 &  27.77245 &  -0.24299  \\
        57182.583973 &  11.22922 &   0.03476 &  27.98556 &  -0.23481  \\
        57183.570990 &  16.72535 &   0.03284 &  27.84478 &  -0.14822  \\
        57184.619044 &  13.51188 &   0.03994 &  28.02373 &  0.01529   \\
        57199.571858 &  19.70855 &   0.04228 &  27.54479 &  -0.61449  \\
        57202.552044 &  10.48449 &   0.03984 &  28.13346 &  -0.40772  \\
        57203.556020 &  18.14702 &   0.04130 &  27.74655 &  -0.40814  \\
        57204.548030 &  12.83536 &   0.03205 &  28.24409 &  -0.43471  \\
        57486.697998 &  19.76909 &   0.03492 &  27.99402 &  -0.27572  \\
        57487.645330 &  9.56572  &   0.03248 &  28.20436 &  -0.32989  \\
        58198.758018 &  9.99469  &   0.03287 &  27.88726 &  0.24160   \\
        58199.708869 &  17.65346 &   0.03003 &  27.82458 &  0.08529   \\
        \bottomrule
    \end{tabular*}
    \label{tab:rv_harps}
\end{table}

\begin{table}
    \sisetup{round-mode=places}
    \centering
    \caption{CORALIE (1) radial velocity dataset. Machine readable format is available
    online at CDS.\newline
    $^*$marks data that was excluded from the fit.}
    \begin{tabular*}{\columnwidth}{@{\extracolsep{\fill}}
                    S[table-format=5.5, group-digits=false, 
                      round-precision=5, table-unit-alignment = center]
                    S[table-format=2.3,
                      round-precision=3, table-unit-alignment = center]
                    S[table-format=1.3,
                      round-precision=3, table-unit-alignment = center]
                    S[table-format=2.3,
                      round-precision=3, table-unit-alignment = center]
                    S[table-format=-1.3,
                      round-precision=3, table-unit-alignment = center]}
        \toprule
        \toprule
        {BJD$_\mathrm{UTC}$} & {RV} & {$\sigma$} & {FWHM} & {BIS} \\
         & {\SI{}{\kilo\metre\per\second}} &
        {\SI{}{\kilo\metre\per\second}} 
        & {\SI{}{\kilo\metre\per\second}} &
        {\SI{}{\kilo\metre\per\second}} \\
        \midrule
        56449.553799 &  9.66479  &  0.07635 &  27.71726 & 0.39581  \\  
        56684.776572 &  19.97341 &  0.08458 &  28.16283 & -0.18934 \\  
        56687.719393 &  12.26985 &  0.08186 &  27.55700 & -0.23831 \\
        56690.804858 &  13.74564 &  0.08728 &  28.12772 & 0.13317  \\
        56692.786749 &  10.85002 &  0.07247 &  27.75353 & -0.01915 \\
        56693.785368 &  19.51221 &  0.07121 &  27.96600 & -0.04120 \\
        56694.751460 &  9.43889  &  0.07732 &  27.92017 & -0.20278 \\
        56696.650910 &  11.26911 &  0.07950 &  28.07056 & -0.37291 \\
        56697.709981 &  17.87517 &  0.07976 &  27.92011 & -0.31469 \\
        56714.747621 &  9.53700  &  0.10260 &  27.63444 & -0.81799 \\
        56718.738063 &  11.08132 &  0.07096 &  28.14193 & -0.27901 \\
        56722.748140 &  17.08727 &  0.07493 &  28.02565 & -0.30170 \\
        56726.607359 &  19.62246 &  0.08759 &  28.05893 & 0.27405  \\
        56739.667157 &  17.83755 &  0.06912 &  28.30868 & -0.37087 \\
        56740.828853 &  11.33148 &  0.08085 &  28.03725 & 0.18074  \\
        56743.729435 &  12.86017 &  0.08500 &  27.87329 & -0.47992 \\
        56748.718570 &  19.68182 &  0.07960 &  28.05396 & 0.58332  \\
        56773.547138 &  17.30455 &  0.09400 &  27.93314 & -0.50840 \\
        56809.616019 &  9.78716  &  0.08697 &  27.88691 & -0.28205 \\
        56810.576861 &  19.97365 &  0.09314 &  27.82375 & 0.03207  \\
        56811.626760 &  10.30808 &  0.08987 &  27.63040 & -0.42166 \\
        56833.553624 &  11.87092 &  0.08970 &  27.65532 & -0.32040 \\
        56837.523574 &  17.99888 &  0.10351 &  28.34447 & 0.25509  \\
        56878.471921 &  13.13879 &  0.10648 &  28.82110 & -0.02549 \\
        56879.472253 &  18.55910 &  0.11714 &  27.56328 & 0.27840  \\
        56880.475420$^*$ &  10.58605 &  0.10760 &  29.58610 & 0.46275  \\
        \bottomrule
    \end{tabular*}
    \label{tab:rv_coralie1}
\end{table}

\begin{table}
    \sisetup{round-mode=places}
    \centering
    \caption{CORALIE (2) radial velocity dataset. Machine readable format is available
    online at CDS.}
    \begin{tabular*}{\columnwidth}{@{\extracolsep{\fill}}
                    S[table-format=5.5, group-digits=false, 
                      round-precision=5, table-unit-alignment = center]
                    S[table-format=2.3,
                      round-precision=3, table-unit-alignment = center]
                    S[table-format=1.3,
                      round-precision=3, table-unit-alignment = center]
                    S[table-format=2.3,
                      round-precision=3, table-unit-alignment = center]
                    S[table-format=-1.3,
                      round-precision=3, table-unit-alignment = center]}
        \toprule
        \toprule
        {BJD$_\mathrm{UTC}$} & {RV} & {$\sigma$} & {FWHM} & {BIS} \\
         & {\SI{}{\kilo\metre\per\second}} &
        {\SI{}{\kilo\metre\per\second}} 
        & {\SI{}{\kilo\metre\per\second}} &
        {\SI{}{\kilo\metre\per\second}} \\
        \midrule
        56998.827826  & 17.47009  & 0.12933  & 27.08932 & -0.39700 \\
        57010.832418  & 10.57601  & 0.09756  & 28.24718 & -0.27460 \\
        57015.847897  & 18.50726  & 0.08972  & 28.09100 & -0.49615 \\
        57023.751810  & 9.79495   & 0.10472  & 28.25546 & 0.04621  \\
        57026.761266  & 17.05405  & 0.09548  & 27.88570 & -0.35298 \\
        57068.721936  & 16.92315  & 0.08204  & 27.84734 & -0.23294 \\
        57079.758041  & 16.58560  & 0.09324  & 27.88199 & -0.24752 \\
        57081.729334  & 13.31412  & 0.09905  & 27.79138 & -0.20482 \\
        57119.490788  & 16.66207  & 0.12742  & 27.87915 & -0.93028 \\
        57121.524136  & 13.88072  & 0.12961  & 27.91771 & -0.24793 \\
        57138.722837  & 9.73737   & 0.11964  & 28.15435 & -0.85951 \\
        57188.507306  & 19.25630  & 0.14644  & 28.72389 & -0.47192 \\
        57370.812711  & 10.00891  & 0.13534  & 28.44965 & 0.13814  \\
        57371.843190  & 19.65781  & 0.12413  & 28.39575 & -0.22403 \\
        57422.738422  & 18.95753  & 0.09442  & 28.50348 & -0.36238 \\
        57423.721681  & 9.50032   & 0.08920  & 27.99541 & -0.06870 \\
        57458.648308  & 11.57403  & 0.09403  & 28.15161 & -0.40348 \\
        57477.536707  & 18.74728  & 0.10241  & 28.08040 & 0.08369  \\
        57560.561020  & 9.27320   & 0.09932  & 28.36695 & 0.08509  \\
        57569.464267  & 9.35399   & 0.11077  & 28.10119 & 0.12780  \\
        57590.489717  & 19.78442  & 0.10282  & 27.71082 & 0.09967  \\
        57716.848800  & 15.01008  & 0.14879  & 27.54259 & -1.16172 \\  
        \bottomrule
    \end{tabular*}
    \label{tab:rv_coralie2}
\end{table}

\section{Photometric information}
\label{appendix:photometric_info}
\begin{table*}
    \sisetup{round-mode=places}
    \centering
    \caption{Light curve information for WASP-128.}
    \begin{tabular}{
                    c
                    c
                    c
                    S[table-format=2.0, table-unit-alignment = center]
                    S[table-format=3.0, table-unit-alignment = center]
                    l}
        \toprule
        \toprule
        Date & Instrument & Filter & $t_\mathrm{exp}$ & $N$ & 
        Baseline function \\
        & & & {s} & & \\
        \midrule
        2013-05-31 & TRAPPIST     & Sloan $z'$ & 8  & 705   & $p(t^2)$  \\
        2014-02-11 & TRAPPIST     & Sloan $z'$ & 11 & 832   & $p(t^2)$ + MF  \\
        2014-02-20 & TRAPPIST     & Sloan $z'$ & 11 & 1014  & $p(t^2)$  \\
        2014-03-03 & TRAPPIST     & Sloan $z'$ & 11 & 860   & $p(t^2)$  \\
        2014-03-03 & \emph{Euler} & Gunn $r'$ & 60 & 215  & $p(t^2)$ + MF  \\
        2014-03-25 & \emph{Euler} & Gunn $r'$ & 75 & 136  & $p(t^2 + \mathrm{sky}^1)$  \\
        2014-05-26 & \emph{Euler} & Gunn $r'$ & 60 & 203  & $p(t^2)$  \\
        2015-05-10 & TRAPPIST     & Sloan $z'$ & 8  & 969   & $p(t^2)$ + MF  \\
        \bottomrule
    \end{tabular}
    \label{tab:photometry}
\end{table*}

\bsp	
\label{lastpage}
\end{document}